\documentstyle[prb,aps,psfig]{revtex}
\begin{document}
\draft
\title{The Band Gap in Silicon Nanocrystallites}
\author{V. Ranjan$^1$, Manish Kapoor$^2$ and Vijay A. Singh$^1$}
\address{ $^1$Physics Department, I.I.T.-Kanpur, U.P. 208016,
  INDIA   and \\ $^2$D.A.V. College, Kanpur, U.P. 208001, INDIA }
\date{March 2001}
\maketitle
\bibliographystyle{prsty}

\begin{abstract}
The gap in semiconductor nanocrystallites has been extensively studied
both theoretically  and experimentally over  the last two  decades. We
have  compared a  recent ``state-of-the-art''  theoretical calculation
with a recent ``state-of-the-art'' experimental observation of the gap
in  Si  nanocrystallite. We  find  that  the  two are  in  substantial
disagreement, with  the disagreement being more  pronounced at smaller
sizes.   Theoretical   calculations   appear  to   over-estimate   the
gap.  Recognizing  that  the   experimental  observations  are  for  a
distribution of crystallite sizes, we proffer a phenomenological model
to reconcile the  theory with the experiment. We  suggest that similar
considerations  must  dictate   comparisons  between  the  theory  and
experiment vis-a-vis  other properties  such as radiative  rate, decay
constant, absorption coefficient, etc.

\end{abstract}

\section{Introduction}
\label{s:intro}
Semiconductor nanocrystallites,  more popularly known  as quantum dots
(QDs), have been extensively studied  over the past decade and a half.
The system  is interesting  from the point  of view of  basic physics,
with   the  carriers   being   confined  to   an  essentially   ``zero
dimensional'' structure.  The efficient luminescence  observed in some
of   these   crystallites   makes   them  promising   candidates   for
opto-electronic devices. Further,  the inexorable drive towards device
miniaturization makes them technologically significant.

The earliest  theoretical works in  this field were reported  in early
1980's  \cite{efro82,brus83}.  For over  a decade  after, most  of the
theoretical works  reported were based on effective  mass theory (EMT)
and  tight-binding semi-empirical  approaches.  These  works predicted
the experimentally observed trends  for quantum confinement, i.e.  the
inverse dependence  of enhanced band gap on  the nanocrystallite size.
These   calculations   were,   however,   performed   for   a   single
nanocrystallite, whereas  experiments are performed on  an ensemble of
QDs  of varying  sizes.   We have earlier pointed out that  the
existence of  an ensemble of QDs  of varying sizes must  be taken into
account  in any  theoretical  formulation \cite{john94,ranj98,kapo99}.
Though  improved  theoretical calculations  were  pursued later,  this
aspect has largely been ignored.

In  a  recent work  Ogut  \textit{et  al.}   \cite{ogut97} reported  a
``state-of-the-art''   theoretical   calculations   based   on   first
principles.  This work is claimed to be in excellent agreement with an
early experimental work \cite{furu88} and is claimed to be superior to
semi-empirical  calculations.   We compared  their  calculations to  a
later ``state-of-the-art''  experimental results of  Buuren \textit{et
al.}   \cite{buur98}.  We found  that there  was a  large disagreement
between    the    two.     This    comparison    is    presented    in
Sec.~\ref{s:compare}.    In    Sec.~\ref{s:phenom},   we   proffer   a
phenomenological  scheme for  this underestimation  of the
band gap and suggest a  possible reconciliation between the theory and
experiment. Conclusions are presented in Sec.~\ref{s:conc}.

\section{Comparison : Theory vs. Experiment}
\label{s:compare}
In  a  Letter Ogut  \textit{et  al.}   \cite{ogut97}  have employed  a
carefully  argued ab-initio methodology  to obtain  the size-dependent
optical gap of Si QDs with sizes ranging from 1 to 3 nm. These authors
choose   to   compare  their   calculations   with  the   experimental
observations of  Furukawa and Miyasato \cite{furu88}.   A little later
\cite{buur98}  Buuren  \textit{et  al.}   reported state  of  the  art
measurements of the  band edges of Si quantum  dots (QDs) with average
diameters ranging from  1 to 5 nm.  By  adding the measured conduction
band (CB) and valence band (VB) shifts to the band gap of bulk Si they
obtained the  band gap of  the Si QDs.   Recent PL and  extended X-ray
absorption data for oxygen terminated silicon nanocrystallites of size
less than 4 nm, were  found to match these observations \cite{schu95}.
These   band  gaps   are  smaller   than  most   reported  theoretical
calculations.   This fact  was noted  by Buuren  \textit{et  al.}  who
choose to  compare their observations with older  calculations by Wang
and   Zunger  \cite{wang96}.    We  also   note  that   several  first
principles/LDA calculations  on nanocrystalline forms of  Si have been
reported  in  the  past  \cite{read92}. It  appears  that  theoretical
calculations over-estimate the gap.

A  comparison  of  the ``state-of-the-art''  theoretical  calculations
\cite{ogut97} with  the ``state-of-the-art'' experimental observations
\cite{buur98}  reveals that  the two  are in  substantial disagreement
with    each    other.     This    comparison    is    presented    in
Fig.~\ref{f:fig1}. In fact the  Ogut \textit{et al.}  calculation does
worse than Wang and Zunger  (not shown in Fig.~\ref{f:fig1}) and other
semi-empirical theoretical calculations.  Perhaps the only commonality
between the theory and experiment  depicted in the figure maybe stated
in terms  of a bland quantum  confinement (QC) dictum:  ``the band gap
increases as the size decreases''.

The disagreement between theory  and experiment is enhanced at smaller
dot sizes. Were we  to make a constant upward shift of  0.72 eV to the
data  of  Buuren  \textit{et  al.}   such that  they  match  with  the
theoretical calculation  at 3.5 nm  then the increasing  divergence at
smaller sizes is clearly manifested.  This is depicted in the inset of
Fig.~\ref{f:fig1}.   On   the  other   hand,  we  may   translate  the
experimental data horizontally by 1.75  nm and force an agreement with
the  (extrapolated) theoretical calculations.   This would  imply that
the QD  sizes have been seriously underestimated  by Buuren \textit{et
al.}.   While the  latter  do  not discount  the  possibility of  some
underestimation, a 1.75 nm error is unlikely.

Ab-initio  calculations  are  computationally  demanding at  large  QD
sizes.  As  we have  pointed out the  disagreement between  theory and
experiment is pronounced at smaller  dot sizes.  It should be possible
to   re-examine  or   repeat   the  ab-initio   calculation  in   this
computationally feasible intermediate regime.

\section{Phenomenology}
\label{s:phenom}
The  observation of  visible photoluminescence  (PL) in  a  variety of
semiconductor  nanocrystallites has  fueled a  large body  of research
work             in             the            past             decade
\cite{john94,ogut97,furu88,bert97,yori97,datt99}.  PL has acquired the
role   of   central   characterizing   tool  in   this   field.    The
photoluminescence  spectra  from such  systems  are  broad, and  often
asymmetric  about  the  peak energy.   The  growth  of  the QDs  is  a
stochastic process.  In an earlier  work we have argued that one needs
to consider the distribution of  crystallite sizes to compare with the
experimental spectral lineshape \cite{john94,ranj98}.  A Gaussian size
distribution   was   used   in   those   works.    For   semiconductor
nanostructures,  the  log-normal  size distribution  has  considerable
experimental      \cite{buur98,kane93,zhan95}      and     theoretical
\cite{kapo99,yori97,datt99}   support.   Specifically,   Yorikawa  and
Muramatsu \cite{yori97}  presented an explanation of  PL spectra based
on  the  log-normal  distribution  of  porous silicon  QD  size.   The
experimental  work reported by  Buuren \textit{et  al.}  also  shows a
log-normal distribution.

We  consider the log-normal  size distribution,  $P(d)$, of  diameter $d$
centered around a mean diameter $d_0$,
\begin{eqnarray}
P(d) & = & \frac{1}{\sqrt{2 \pi} \sigma d} \exp \left[-
\frac{\left(\ln d - \mu \right)^2}{2 \sigma^2} \right] \label{eq:log} \\
d_m & = & \exp (\mu - \sigma^2) \label{eq:dmax} \\
d_0 & = & \exp \left(\frac{\sigma^2}{2} + \mu \right) \label{eq:dmean} 
\end{eqnarray}
where  $d_m$ is  the  dot size  for  which the  maxima  occurs in  the
log-normal  distribution and  \{$\mu,\sigma$\} are  some characteristic
constants.

The number of electrons in a dot of diameter $d$ participating in a PL
process is  proportional to $d^3$. Thus,  for an ensemble  of QDs, the
probability distribution of electrons  participating in the PL process
is
\begin{eqnarray}
P_e(d) & = & \frac{1}{\sqrt{2 \pi} \sigma d} bd^3\exp \left[-
\frac{\left(\ln d - \mu \right)^2}{2 \sigma^2} \right] \label{eq:loge}
\end{eqnarray}
where $b$ is a suitable normalization constant.

In general, the  optical band gap is attributed  to the energy upshift
of the electron and yield
\begin{eqnarray}
E & = & E_{\infty} + \frac{C}{d^n}  \label{eq:up}
\end {eqnarray}
where $E$  is the enhanced gap,  $E_{\infty}$ is the  bulk silicon gap
(1.17 eV),  $C$ is an appropriately  dimensioned constant and $n$  is the
exponent \cite{john94,ranj98} with $n \in [1,2]$.

Hence the energy upshift, $\Delta E$, due to confinement in QD is
\begin{eqnarray}
\Delta E & = & \frac{C}{d^n} \label{eq:shift} \\
\Delta E_0 & = & \frac{C}{d_0^n} \label{eq:mean}
\end{eqnarray}
where $\Delta  E_0$ is a  mean upshift, which  is related to  the mean
diameter $d_0$ of QD.

Now,   convoluting  the   upshift  (eqn.~(\ref{eq:shift}))   with  the
log-normal size distribution (eqn.~(\ref{eq:loge}))
\begin{eqnarray}
P(\Delta E) & = & \int_0^{\infty} \delta (\Delta E - \frac{C}{d^n}) \frac{d^2}{\sqrt{2
\pi} \sigma} \exp \left[-
\frac{\left(\ln d - \mu \right)^2}{2 \sigma^2} \right] d(d) 
\end{eqnarray} 
The above  integral can be easily  solved using the  property of Dirac
delta function, which yields
\begin{eqnarray}
P(\Delta E) & = & \frac{b}{\sqrt{2\pi}\sigma n C}\left(\frac{C}{\Delta E}
\right)^{(3+n)/n}\exp\left[-\frac{\{(1/n)\log(C/\Delta E) - \mu\}^2}{2 
\sigma^2} \right]  \label{eq:dirac}
\end{eqnarray}
For the PL peak, we equate the derivative of the eqn.~(\ref{eq:dirac})
to zero.  The shift in the PL peak position, $\Delta E_p$, is given by
\begin{eqnarray}
\Delta E_p & = & C \exp \left[- \{ (3 + n) \sigma^2 + \mu \} n \right]
\\
         & = & \frac{C}{d_0^n} \left( \frac{d_m}{d_0} \right)^{n [2n +
5]/3} \label{eq:peak}
\end{eqnarray}
We employ eqn.~(\ref{eq:peak}) in eqn.~(\ref{eq:up}) to get
\begin{eqnarray}
E & = & E_{\infty} + \Delta E_p \\
  & = & E_{\infty} + \frac{C}{d_0^n} \left( \frac{d_m}{d_0}
  \right)^{n[2n + 5]/3} \label{eq:final}
\end{eqnarray}
Eq.~(\ref{eq:final})  represents the  final form  for  estimating a more
realistic size dependence of the optical  gap. The exponent $n = $1.22 and
$C =  $ 3.9 in  appropriate units, for  the results of  Ogut \textit{et
al.}    shown  in   fig.~(\ref{f:fig1}).    Kanemitsu  \textit{et~al.}
\cite{kane93}   obtained  the   size  distribution   of   oxidized  Si
nanocrystallites  produced  by laser  breakdown  of  silane gas.   The
optimum log-normal  fit on these experimental data  yield $d_m/d_0 \in
[0.7,0.95]$.  We have used three values  0.7, 0.8 and 0.9 of the ratio
$d_m/d_0$  in eqn.~(\ref{eq:final})  and depicted  them  alongside the
data  from  Ogut  \textit{et  al.}   and Buuren  \textit{et  al.}   in
fig.~(\ref{f:fig2}). We  see that our  calculation with the  value 0.7
brings Ogut \textit{et al.}'s result in complete agreement with Buuren
\textit{et  al.}'s result.  Further our  phenomenological model  produces more
pronounced downshift at  smaller sizes. This is once again in  agreement with the
experimental trend.  The asymmetric  distribution of the dot size with
stretched  tailing  towards  larger  QDs  is  responsible  for  the  lower
$d_m/d_0$  ratio, though  the skewness  may  not be  as pronounced  as
$d_m/d_0$ =  0.7.  Thus,  for more asymmetric  log-normal distribution
one obtains more prominent downshift at smaller sizes.

\section{Conclusion}
\label{s:conc}
We  find that a  recent ``state-of-the-art''  theoretical calculation
\cite{ogut97} for optical gap is in large disagreement with a closely
following  ``state-of-the-art'' experimental work  \cite{buur98}.  The
comparison  is  even  worse   for  smaller  crystallite  sizes.   The
experimental   work  reported  a   log-normal  size   distribution  for
nanocrystallite samples, whereas the theoretical work was performed on
a single crystallite.  This theoretical calculation showed a very good
agreement  with   a  decade  old  experimental   work,  whereas  other
semi-empirical  works were  shown  to  be in  poor  agreement. To  the
contrary,  we  find  that  the  semi-empirical works  were  in  better
agreement with Buuren \textit{et al.}'s experimental work.

Ab-initio  calculations  are  computationally  demanding at  large  QD
sizes.  As we  have pointed  out the  disagreement between  theory and
experiment is pronounced at smaller  dot sizes.  It should be possible
to re-examine or repeat the  ab-initio calculation in this computationally
feasible intermediate regime.

We   have   used   the   methodology  discussed   in   earlier   works
\cite{john94,ranj98,kapo99} with  a log-normal distribution.   We have
shown that Ogut  \textit{et al.}'s results can be  in better agreement
with experimental works if  size distribution is explicitly taken into
account.  In  our formulation the  downshift of optical gap  is larger
for smaller crystallite sizes, making this exercise even more relevant
to the present case.

We caution  however that the  disagreement between the theory  and the
experiment may not be solely  related to a distribution of crystallite
sizes.  Other  factors could  be  crucial.  These  are :  (i)  partial
passivation, (ii) passivation with  species other than hydrogen, (iii)
surface  reconstruction, (iv) flattening  of the  dot, and  (v) size
underestimation by  the microscopic techniques. Some  of these factors
have been mentioned by Buuren \textit{et al.}  \cite{buur98}.

There has been a growing realization that the existence of an ensemble
of QDs of varying sizes must be taken into account in order to explain
experimental                                               observations
\cite{john94,ranj98,kapo99,bert97,yori97,datt99,josh95,fu97}. Tentative
attempts  have  been made  to  understand  the  radiative rate,  decay
constant, absorption data, etc. based on these considerations. Perhaps
any  theoretical calculation  on  a single  nanocrystallite should  be
supplemented with  effective size  averaging before a  comparison with
experiment  is  made.  

\section*{Acknowledgement}  
This work  was  supported  by the  Department  of Atomic  Energy
through the Board of Research in Nuclear Sciences, India.

\newpage
\section*{Figure Caption}

\begin{figure}
\caption{The  figure  compares  the ``state-of-the-art''  experimental
band  gap  obtained  by  Buuren  \textit{et al.}   [Ref.~8]  with  the
``state-of-the-art'' theoretical one  obtained by Ogut \textit{et al.}
[Ref.~6]. The  two are in  considerable disagreement. In the  inset we
have up-shifted the data from Ref.~8 by 0.72 eV to ensure an agreement
with Ref.~6  at 3.5 nm size.   The discrepancy between  the theory and
the experiment appears pronounced at smaller sizes. The dashed line is
our fit to  the calculations of Ref.~6 ($E_g(d)  = 1.1 + c/d^{1.22}$).
The `$+$' symbols are data  arrived at by using a relationship between
CB and VB edges by Buuren \textit{et al.} [see Ref.~8].}
\label{f:fig1}
\end{figure}

\begin{figure}
\caption{The  figure depicts the  theoretical data  of Ref.~6  and the
experimental  data  of  Ref.~8.   It   also  shows  the  band  gap  of
crystallites if a log-normal size distribution is incorporated in Ogut
\textit{et  al.}'s calculation.   Of the  four lines  depicted  in the
figure the lowest one (dashed line) and the subsequent two (dotted and
dashed lines) corresponds to log-normal distributions with : $d_m/d_0$
= 0.7, 0.8 and 0.9 respectively. The topmost dotted line is the fit to
the data of Ref~6 as mentioned  in Fig~1 and the text.  We notice that
the  downshift of  the  band  gap is  larger  for smaller  crystallite
sizes. We also notice that $d_m/d_0$ =  0.7 shows excellent agreement
with the experimental results of Buuren \textit{et al.} [Ref.~8].}
\label{f:fig2}
\end{figure}

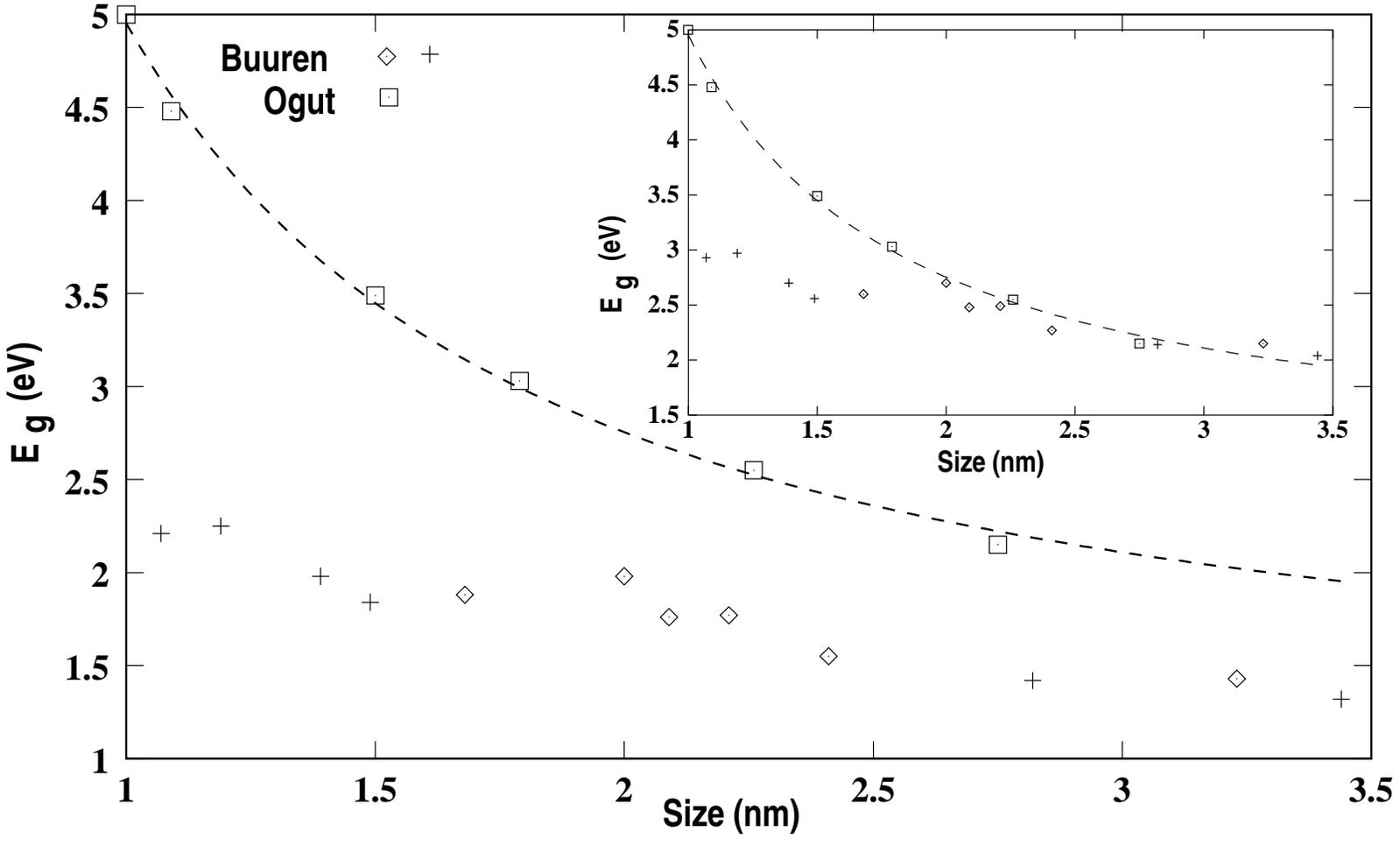
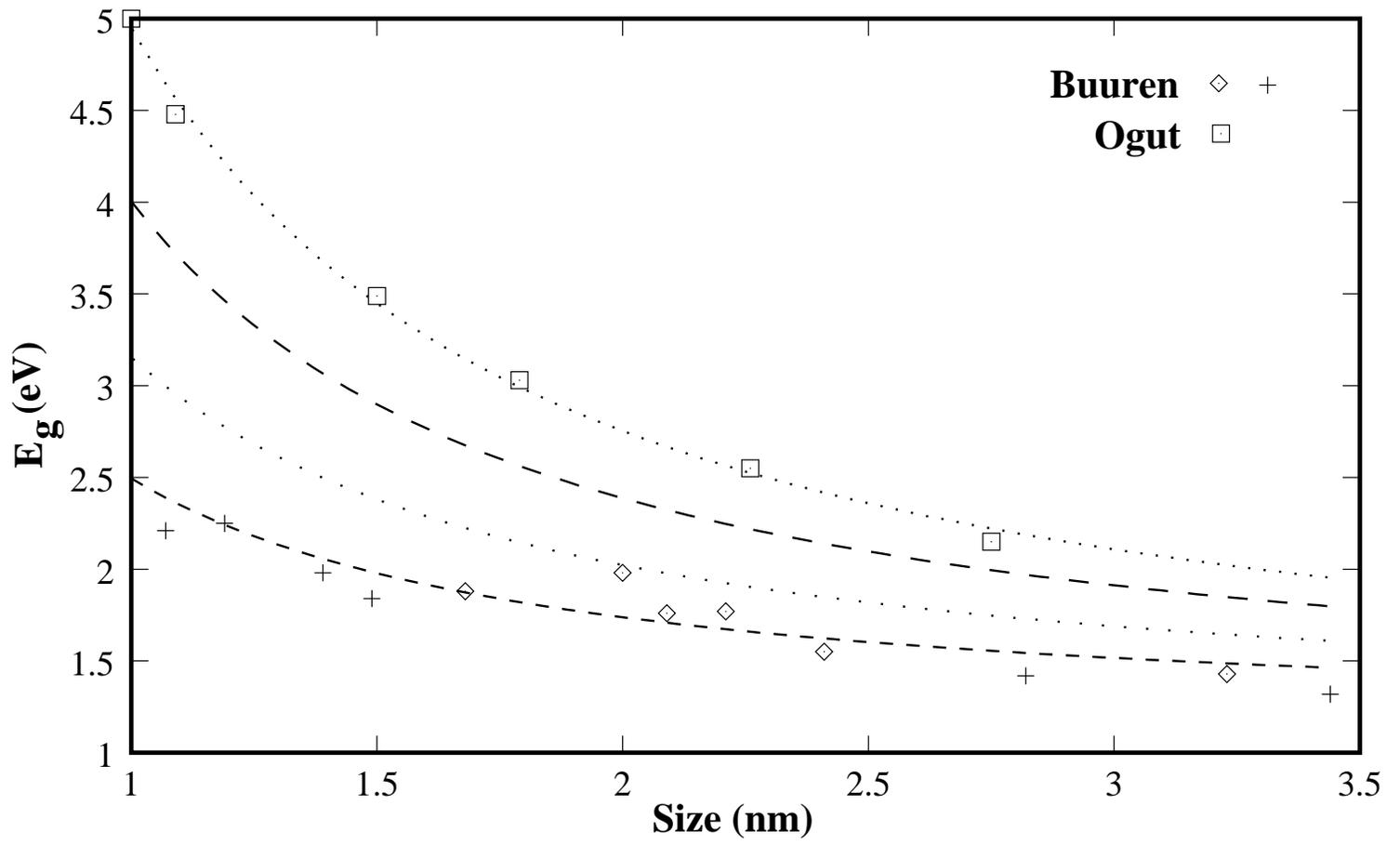
\end{document}